\documentclass[12pt,preprint]{aastex} 

\shorttitle{$^6$Li in the atmosphere of GJ~117}
\shortauthors{D. Christian et~al.}
\begin{document}

\title{$^6$Li in the atmosphere of GJ 117 Revisited}

\author{D.J. Christian\footnote{Current address: Department of Physics and Astronomy,
California State University, Northridge, 18111 Nordhoff Street, Northridge, CA 9
1330}, M. Mathioudakis,}
\affil{Astrophysics Research Centre, Queen's University Belfast,
	Belfast, BT7 1NN, Northern Ireland, U.K.}
\and
\author{D. Jevremovi\'c }
\affil{Astronomical Observatory, Volgina 7, 11160 Belgrade, Serbia and Montenegro }



\begin{abstract}
Detection of $^6$Li has been shown for energetic
solar events, one chromospherically active binary, and several dwarf
halo stars.
We had previously found a $\frac{^{6}Li}{^{7}Li}$ = 0.03$\pm0.01$ 
for active K dwarf GJ~117 using VLT UVES observations. 
Here we present high signal-to-noise ($>$1000) 
high spectral resolution observations taken with the McDonald 
Observatory's 2.7m and echelle spectrometer of GJ~117. 
We have used the solar spectrum and template stars to eliminate possible
blends, such as Ti~I, in the $^6$Li spectral region.
Our new analysis, using an updated PHOENIX model atmosphere finds 
$\frac{^{6}Li}{^{7}Li}$ = 0.05$\pm$0.02. 
Additionally, bisector analysis showed no significant red asymmetries
that would affect the lithium line profile.
No changes above the statistical uncertainties are found
between the VLT and McDonald data.
The amount of $^6$Li derived for GJ~117 is consistent with creation
in spallation reactions on the stellar surface, but
we caution that uncertainties in the continuum level may cause  
additional uncertainty in the $^6$Li fraction.

\end{abstract}


\section{Introduction}
\label{intro}
The destruction of primordial lithium in low-mass stars occurs during 
the pre-main-sequence state and little or no Li is therefore expected 
in these objects when they arrive on the main sequence. 
Studies of open clusters found that the largest lithium abundances
occur for the fastest rotating stars \citep{S93}.  This result is in contradiction
to the expectation that fast rotation enhances the mixing process that
leads to increased lithium depletion, and the rotational history must
play an important role in the mixing process \citep{P97}.    

One of the key assumptions in the models of Li
depletion, is that lithium can not be produced on the surface of late-type
stars and its abundance will only decrease as a function of
time. However, if lithium can be produced on the stellar surface, a
"quasi-equilibrium" abundance would be reached which could depend on
the activity level. Thus, we have been investigating if 
$^6$Li can be formed in active stars by spallation in stellar flares as
a result of their high levels of activity.

The presence of $^6$Li in metal poor halo stars sparked much debate 
in the early 1990s \citep{SLN93}. 
Subsequent observations confirmed the detection of $^6$Li in 
such halo stars and its production was attributed 
to formation by cosmic ray spallation in the ISM \citep{SLN98,HT97}, although
more recent work argues for a pre-Galactic origin \citep{A06}. 

The detection of $^6$Li 
enhancement has been reported during an energetic flare on a 
chromospherically active binary \citep{MR98}, and an active
K dwarf \citep{C05}. There are also indications that this isotope may 
be enhanced in sunspots \citep{G84, R97}. The generation of significant 
$^6$Li in stellar and solar flares has been predicted by several authors 
\citep{C75, W85, L97}.  Such enhanced $^6$Li abundances should also be expected
for stellar objects if extreme energetic conditions are met \citep{ML99}.  
   
In an earlier paper we have shown the detection of $^6$Li in the atmosphere 
of the active K dwarf GJ~117 (=HD~17925) at the 3\% level \citep{C05}. 
Given the significance of this result we have decided to re-observe GJ 117 
at a much higher signal-to-noise than observed so far. Our observations, 
analysis, and models (which are given in more detail in \citet{C05}) are 
presented in \S\ 2. Our results are discussed in \S\ 3 with concluding 
remarks presented in \S\ 4.

\section{Observations \& Data Analysis}
\subsection{Observations}
\label{obs}
\begin{figure}
\includegraphics[scale=0.66, angle=90]{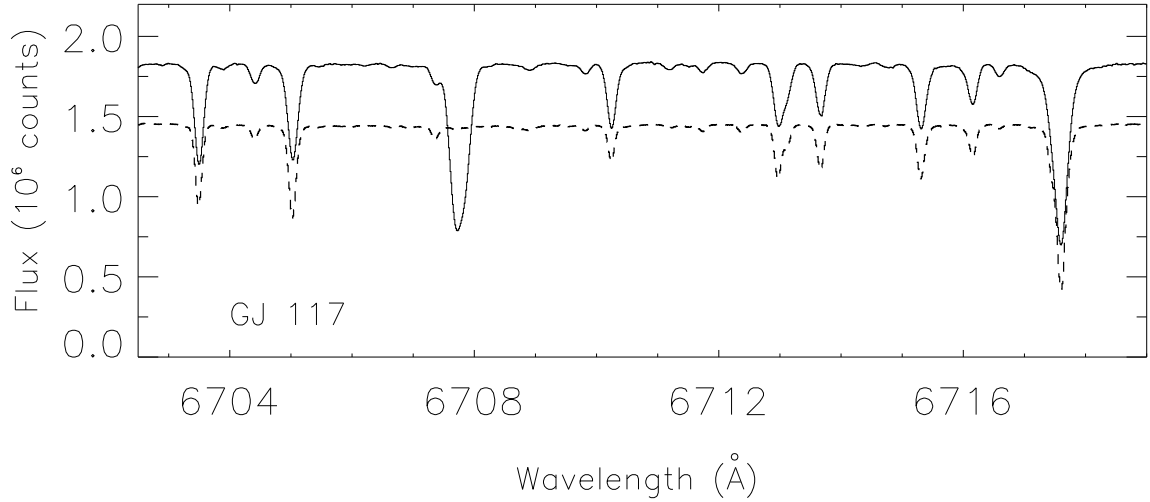}
\caption{McDonald echelle spectra of GJ~117 (solid histogram) compared to the
solar spectrum (dashed line) in the Li I 6707.8\AA\ region. 
The spectra are shown in total counts and 
the solar spectrum is offset by 30\% for ease of presentation.
}
 \label{fig:bigspec}
\end{figure}

Observations of GJ~117 were conducted between 03 -- 09 November 2006 
using 107 inch Harlan J. Smith telescope and echelle spectrograph at
McDonald Observatory. 
The CS21-e1 instrument was used with the E1 grating 
centered at 6708\AA\ and the TK3 2k$\times$2k CCD. 
This set-up provided a resolution of $\approx$130,000.
Multiple exposure of GJ~117 were taken for a combined spectrum with
55 ksec of exposure and a signal-to-noise of over $\approx$1300. 
The data were reduced with standard {\sc iraf} tasks, {\sc imred.ccdred} and 
{\sc twodspec.apextract}, and further analysis carried out using {\sc idl} and  the {\sc starlink} based {\sc dipso} software.
The spectrum of GJ~117 compared to the solar spectrum (taken with the same
set-up) in the Li~I 6708\AA\ region is shown in Figure~\ref{fig:bigspec}.

We also obtained Extreme Ultraviolet Explorer (EUVE) data from the Multimission Archive at Space Telescope.
EUVE observed GJ~117 for nearly 200 ksec in December 1994 with a mean count rate
in the Deep Survey (DS) 100 \AA\ bandpass of 0.09 counts s$^{-1}$. 
We reduced the EUVE data with the standard IRAF {\sc EUV} reduction tasks 
and constructed light curves for the DS data using the {\sc xray.xtiming} 
package as discussed previously in \citet{CM98}.
A log of the GJ~117 observations are listed in Table 1. 

\begin{deluxetable}{llccl}
\tablewidth{0pt}
\tablenum{1}
\tablecaption{OBSERVATION LOG}
\tablehead{
\colhead{Name}
  &\colhead{Observation}
  &\colhead{Exposure}
  &\colhead{Air Mass}
  \\
  \colhead{}
  &\colhead{Date}
  &\colhead{(sec)}
  &\colhead{}
}
\startdata
\multicolumn{4}{c}{McDonald} \\ 
GJ 117\tablenotemark{a} & 2006 Nov 04 & 1800 & 1.44  \\  
& ~~~~~~~~~~~ & 1800 & 1.39  \\  
& ~~~~~~~~~~~ & 1800 & 1.38  \\  
& ~~~~~~~~~~~ & 1800 & 1.39  \\  
& ~~~~~~~~~~~ & 1800 & 1.56  \\  

& 2006 Nov 05 & 1800 & 1.45  \\  
& ~~~~~~~~~~~ & 2500 & 1.39  \\  
& ~~~~~~~~~~~ & 2500 & 1.38  \\  
& ~~~~~~~~~~~ & 2000 & 1.65  \\  

& 2006 Nov 06 & 1800 & 1.49  \\  
& ~~~~~~~~~~~ & 3000 & 1.42  \\  
& ~~~~~~~~~~~ & 3000 & 1.38  \\  

& 2006 Nov 07 & 3000 & 1.39  \\  
& ~~~~~~~~~~~ & 1800 & 1.48  \\  

& 2006 Nov 08 & 1800 & 1.68  \\  
& ~~~~~~~~~~~ & 3000 & 1.52  \\  
& ~~~~~~~~~~~ & 1800 & 1.86  \\  

& 2006 Nov 09 & 1800 & 1.42  \\  
& ~~~~~~~~~~~ & 3000 & 1.38  \\ 
& ~~~~~~~~~~~ & 1800 & 1.46  \\  

  GJ~211\tablenotemark{b}          & 2006 Nov 04 & 900 & 1.13   \\
             &                        & 1800 & 1.65 \\ 
             & 2006 Nov 05 & 900 & 1.10 \\
             &             & 900 & 1.11 \\
             & 2006 Nov 06 & 1200 & 1.15 \\
\\
\multicolumn{4}{c}{EUVE} \\ 
  GJ~117        &1994 Dec 02 & 192\tablenotemark{c}  & ...   \\

\tablenotetext{a}{GJ 117: RA$_{J2000}$= 02 52 32.1; Dec$_{J2000}$= $-$12 46 11.0}
\tablenotetext{b}{Stellar template (see text).} 
\tablenotetext{c}{EUVE observation time in kiloseconds.} 
\enddata
\end{deluxetable}

\begin{figure}
\includegraphics[scale=0.66,angle=90]{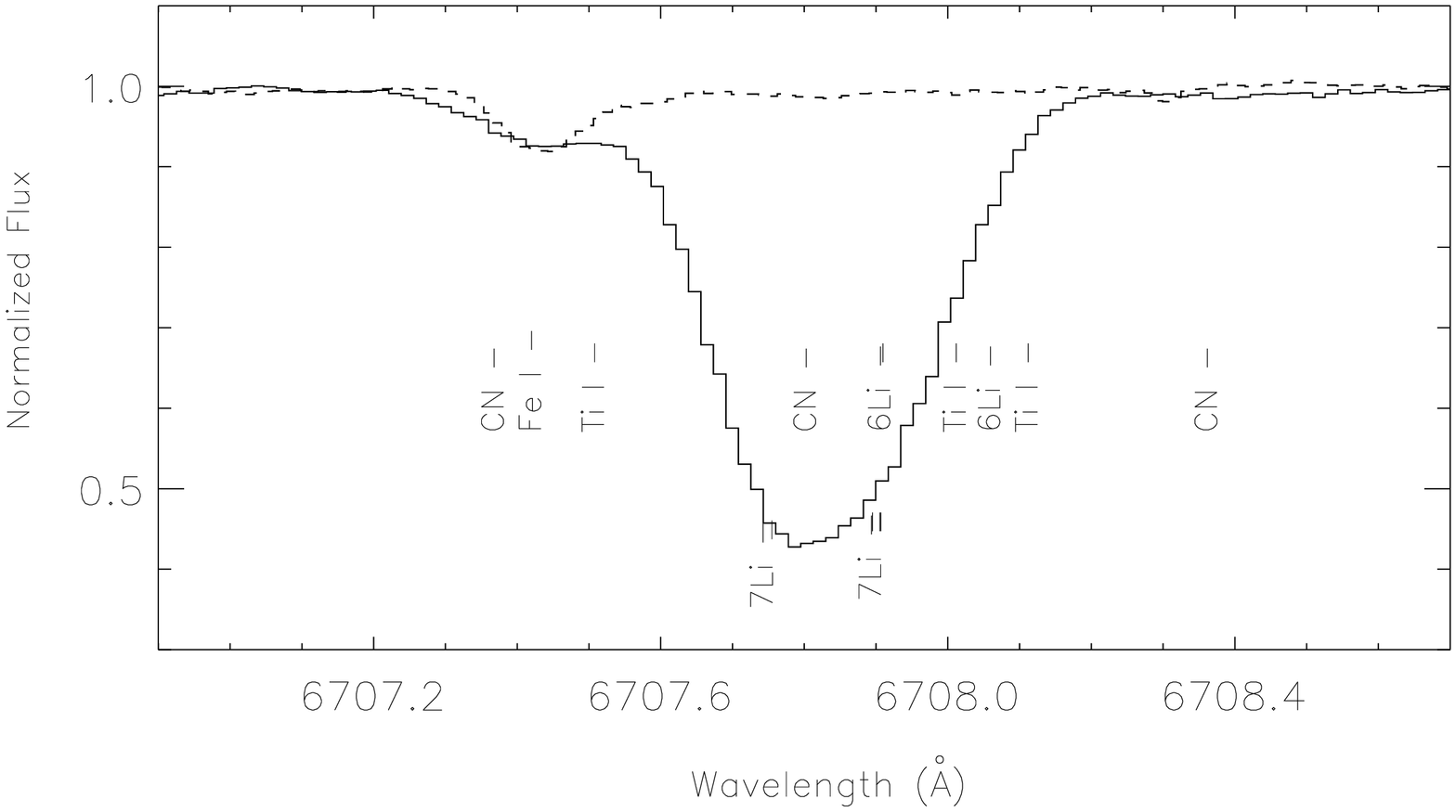}
\caption{
Comparison of the McDonald GJ 117 spectrum (solid histogram) to that of a K2 template star, GJ~211 (dashed line).
}
\label{fig:template}
\end{figure}
\medskip
\subsection{Model Spectra}
\label{models}

$^6$Li is separated from the $^7$Li doublet only by 0.16 \AA\ and has a strength typically less than a few
percent of that feature. In addition to the feature's intrinsic weakness, the situation is further
complicated by blending with nearby lines, such as CN, Fe~I,  and Ti~I \citep{N99, R02}. Contamination by
such lines becomes more prevalent as we move down the main sequence to cooler metal-rich stars.  Template
stars were observed along with GJ~117 to try to resolve the issue of blending. These stars were chosen to
have a log(T) similar to GJ~117, and of the several observed, GJ~211 (HD~37394) also has the closest
metallicity  (Fe/H = 0.12), compared to GJ 117 (Fe/H=0.15; \citet{LH05}). The spectra of GJ~211 and GJ~117
are compared in Figure~\ref{fig:template}.  CN lines near 6707.5 that have been seen to be $\leq$ 1 m\AA\
in the solar spectrum, are also very weak in GJ~211, with an equivalent width upper limit of $\approx$0.5
m\AA. CN should have a negligible effect on the GJ~117 Li I line profile. Additionally there is no 
evidence for a strong Ti~I line near 6708.025 \AA\ with a conservative upper limit of 
$\leq$ 1 m\AA. 
%
 
The general stellar atmosphere code {\sc Phoenix} 
\citep{HS95, AH95} was used to calculate theoretical stellar spectra.  
We have used the {\sc Phoenix} NextGen series LTE atmosphere models 
\citep{haus99} for the effective temperature (log($T_{eff}$)=3.7) and 
gravity (log($g$)=4.6) of GJ~117 \citep{G05, Cin07}. 
We constructed a grid of models from NextGen that sampled 
every 200K in T$_{eff}$ and 0.5dex in log($g$).
Direct opacity sampling \citep{H01} is used within {\sc Phoenix} to
handle any line blanketing, where the  
opacity at each wavelength point is calculated as a sum of opacities
from all the contributing species. 
We added the $^6$Li resonance doublet to the {\sc Phoenix} master line list
\citep{K95} using the wavelengths and $gf$ values of \citet{SLN93}.
A Ti abundance of log(Ti) = 5.22 was chosen from \citet{LH05}. 
We caution that the Ti abundance and $\frac{^{6}Li}{^{7}Li}$ ratio are anti-correlated 
and increasing the Ti abundance decreases the $\frac{^{6}Li}{^{7}Li}$ ratio and conversely,
decreasing the Ti abundance increases the $\frac{^{6}Li}{^{7}Li}$ ratio.

\begin{figure}
\includegraphics[scale=0.66, angle=0]{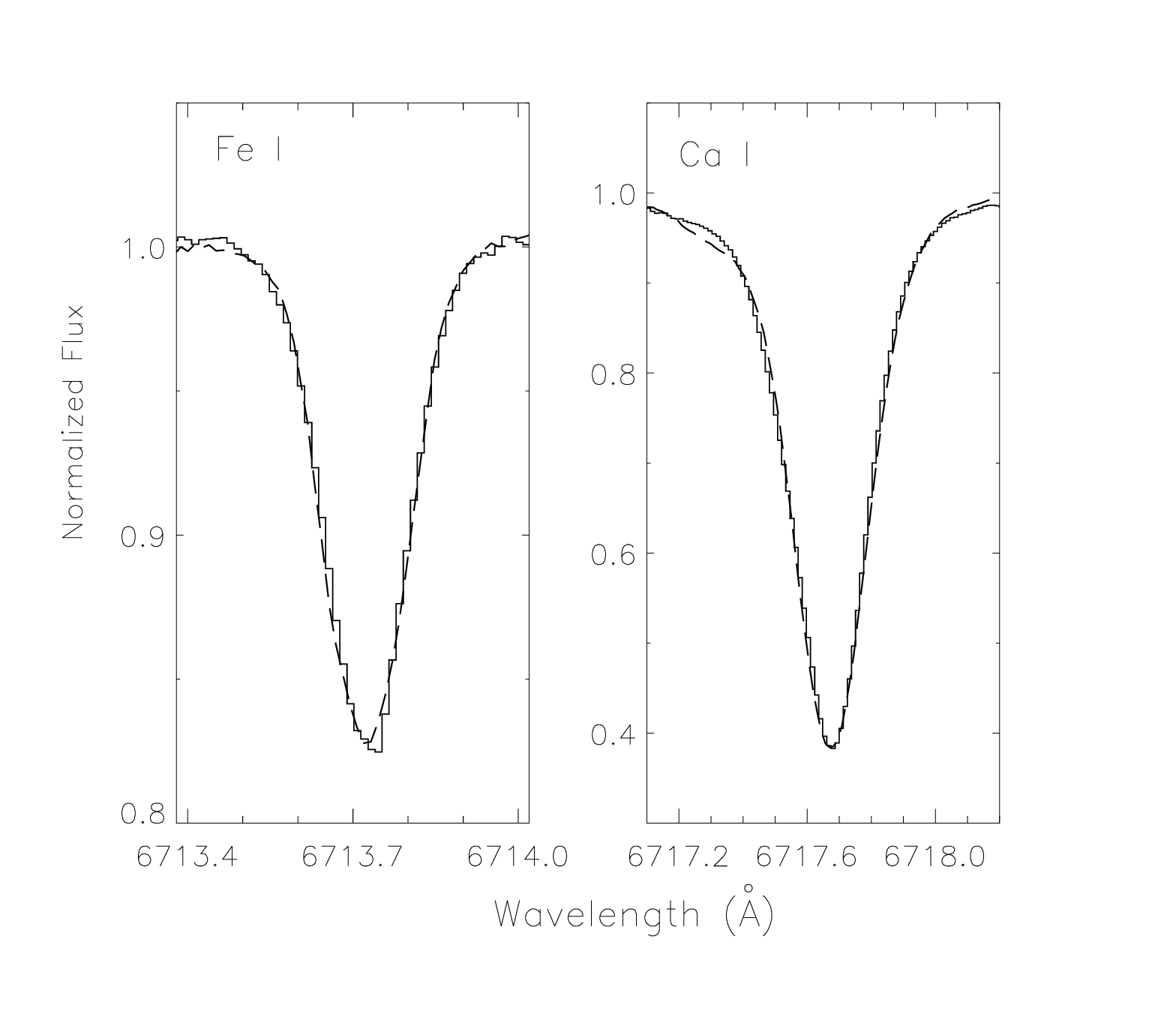}
\caption{Sample synthetic line profiles for: Fe I $\lambda$6713.7 ({\it left}), 
and Ca I $\lambda$6717.68 ({\it right}) lines using the  
derived $v$sini of 5.5 km s$^{-1}$ (see text).
}
\label{fig:profile}
\end{figure}

Model spectra were calculated with a microturbulence 
velocity of $\xi$=1.5 km s$^{-1}$. This value is typical for solar-type stars and is
consistent with the microturbulence given for GJ~117 in \citet{AP04}. 
We derived a $v$sini of GJ~117 of 5.5$\pm$0.2 km s$^{-1}$ using 
$\chi^{2}$ analysis of several lines in the Li~I 6707.8\AA\ region \citep{C05}.
We show model fits for the Fe~I $\lambda$6713.7 and Ca~I at $\lambda$6717.685 
line profiles in Figure~\ref{fig:profile} 
using the derived $v$sini of 5.5 km s$^{-1}$.
 
\citet{R02} have shown that the apparent detection of $^6$Li 
for planet hosting star HD~82943
could be explained by the presence of Ti I lines near 6708.03 and 6708.1 \AA. 
 Although our template spectrum of GJ~211 rules out any strong Ti~I features
in the red-side of the lithium line profile we constructed 
a second set of {\sc Phoenix} models (PHX2) with the 
Ti~I values of \citet{R02} for comparison and completeness. 
We show the line list for important lines in the vicinity of the lithium
line profile and their comparison to \citet{R02} in Table~2.
This second {\sc Phoenix} model (PHX2) has the same values as shown for the
first {\sc Phoenix} model in Table~2, but includes the 
\citet{R02} values for the Ti lines at 6707.752, 6708.025 and 
6708.125 \AA.


\begin{deluxetable}{lcccccc}
\tablewidth{0pt}
\tablenum{2}
\tablecaption{Lines in the vicinity of the Li I 6707.8 \AA\ profile}
\tablehead{
 \colhead{Wavelength}
  &\colhead{Element}
  &\multicolumn{2}{c}{LEP (eV)}
  &\colhead{}
  &\multicolumn{2}{c}{log~gf}
\\
\cline{3-4} \cline{6-7}
\\
  \colhead{\AA}
  &\colhead{}
  &\colhead{R02\tablenotemark{a}}
  &\colhead{PHX\tablenotemark{a}}
  &\colhead{}
  &\colhead{R02}
  &\colhead{PHX}
}
\startdata
6707.3810  & CN    & 1.83 & 1.83 & & $-$2.170 & $-$2.141 \\
6707.4330  & Fe I  & 4.61 & 4.61 & & $-$2.283 & $-$2.283 \\
6707.4500  & Sm II & 0.93 & 0.93 & & $-$1.040 & $-$2.379  \\
6707.4640  & CN    & 0.79 & 0.78 & & $-$3.012 & $-$2.921  \\
6707.5210  & CN    & 2.17 & 2.00 & & $-$1.428 & $-$1.358  \\
6707.5290  & CN    & 0.96 & 0.96 & & $-$1.609 & $-$1.594  \\
6707.5630  & V I   & 2.74 &2.74 & & $-$1.530 & $-$1.995  \\
6707.6440  & Cr I  & 4.21 &4.21 & & $-$2.140 & $-$2.667  \\
6707.7520  & Ti I  & 4.05 &4.77 & & $-$2.654 & $-$2.656  \\
6707.7610\tablenotemark{b} & $^7$Li   & 0.00 & 0.00 & & $-$0.002 & $-$0.009  \\
6707.7710  & Ca I  & 5.80 & 5.79 & & $-$4.015 & $-$4.014  \\
6707.8160  & CN    & 1.21 & 1.21 & & $-$2.317 & $-$1.967  \\
6707.9130\tablenotemark{c} & $^7$Li  & 0.00 &0.00 & & $-$0.807 & $-$0.309  \\
6707.9210 & $^6$Li   & 0.00 &0.00 & & $-$0.002 & $-$0.005  \\
6708.0250  & Ti I   & 1.88 & ... & & $-$2.252 & ...  \\ 
6708.0728  & $^6$Li & 0.00 & 0.00 & & $-$0.303 & $-$0.309 \\
6708.0940  & V I    & 1.22 & 1.22 & & $-$3.113 & $-$3.113  \\ 
6708.1250  & Ti I   & 1.88 & 5.65 & & $-$2.886 & $-$2.801 \\ 
6708.3750  & CN    & 2.10 & 1.89 & & $-$2.252 & $-$1.959 \\ 
\\
\tablenotetext{a}{R02 -- from the line list of \citet{R02}; PHX -- from the PHOENIX line list used in this work.} 
\tablenotetext{b}{Ave of $\lambda\lambda$6707.754 and 6707.766 $^7$Li lines}
\tablenotetext{c}{Ave of $\lambda\lambda$6707.904 and 6707.917 $^7$Li lines}
\enddata
\end{deluxetable} 

%
%

\begin{figure}
\includegraphics[scale=0.66, angle=0]{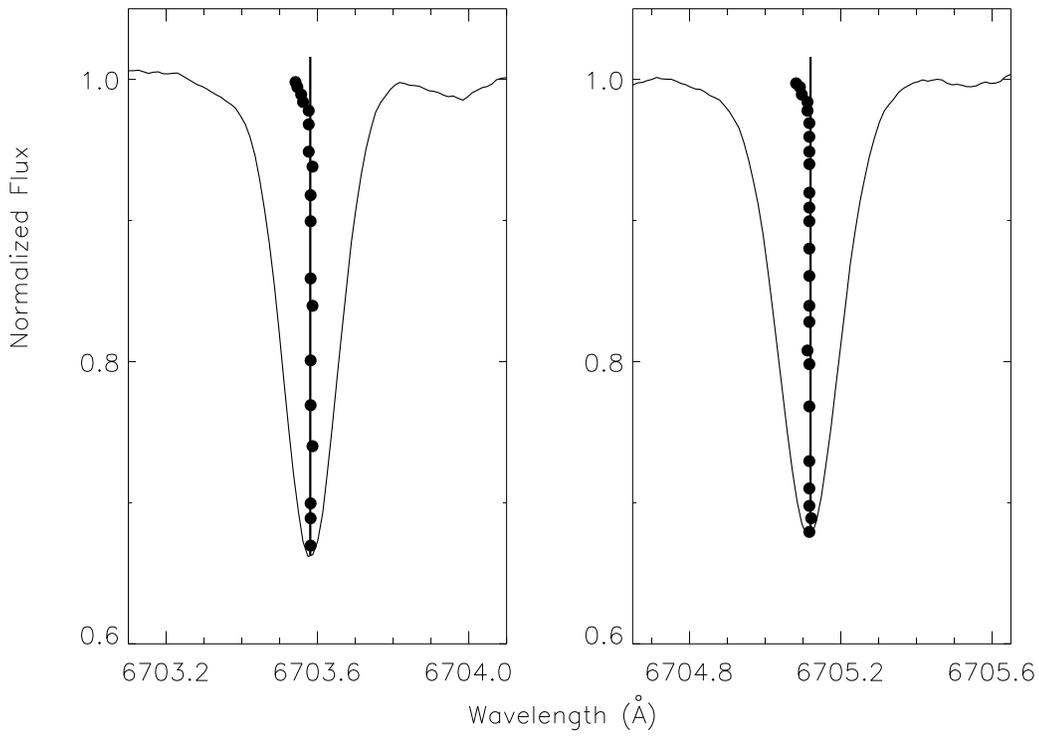}
\caption{Sample line bisectors for 2 Fe I lines ($\lambda$6703.5 \& $\lambda$6705.1) 
in the vicinity of the lithium line profile.
}
\label{fig:bisector}
\end{figure}

  Bisector analysis has been shown to be useful in disentangling 
subtle enhancements in line profiles that may be caused by
blending with weaker lines and indicate other effects in the 
stellar atmosphere, such as granulation. 
In a recent paper, \citet{C07} have used bisector analysis  in observed line profiles of HD 74000 and have
shown that previously un-detected line asymmetries may cause derived lithium fractions to be lower than 
previously reported. We have performed bisector analysis on 3 Fe lines (6703.5, 6705.1, 6713.7\AA) in the
Li 6708\AA\ region to evaluate the effect of asymmetries. 
%
We show the line bisectors for 2 of these Fe lines in Figure~\ref{fig:bisector}.
We find that the line bisectors for these lines on GJ 117 are vertical,  
and show a blue asymmetry near the top of the profile, but show no red-symmetry.
We estimate an upper limit to red-asymmetry of $\approx$0.005\AA\ (0.2 km/s). This is consistent with 
our earlier estimates from VLT UVES spectra of GJ~117. We therefore conclude, 
that there are no significant red asymmetries that may affect the Li line 
profile.
We discuss uncertainties in continuum placement in Section 3.1,
which compares the McDonald and VLT spectra of GJ~117.

\section{Results \& Discussion}
\subsection{The $\frac{^{6}Li}{^{7}Li}$ isotope ratio}

\begin{figure*}
\includegraphics[scale=0.63,angle=90]{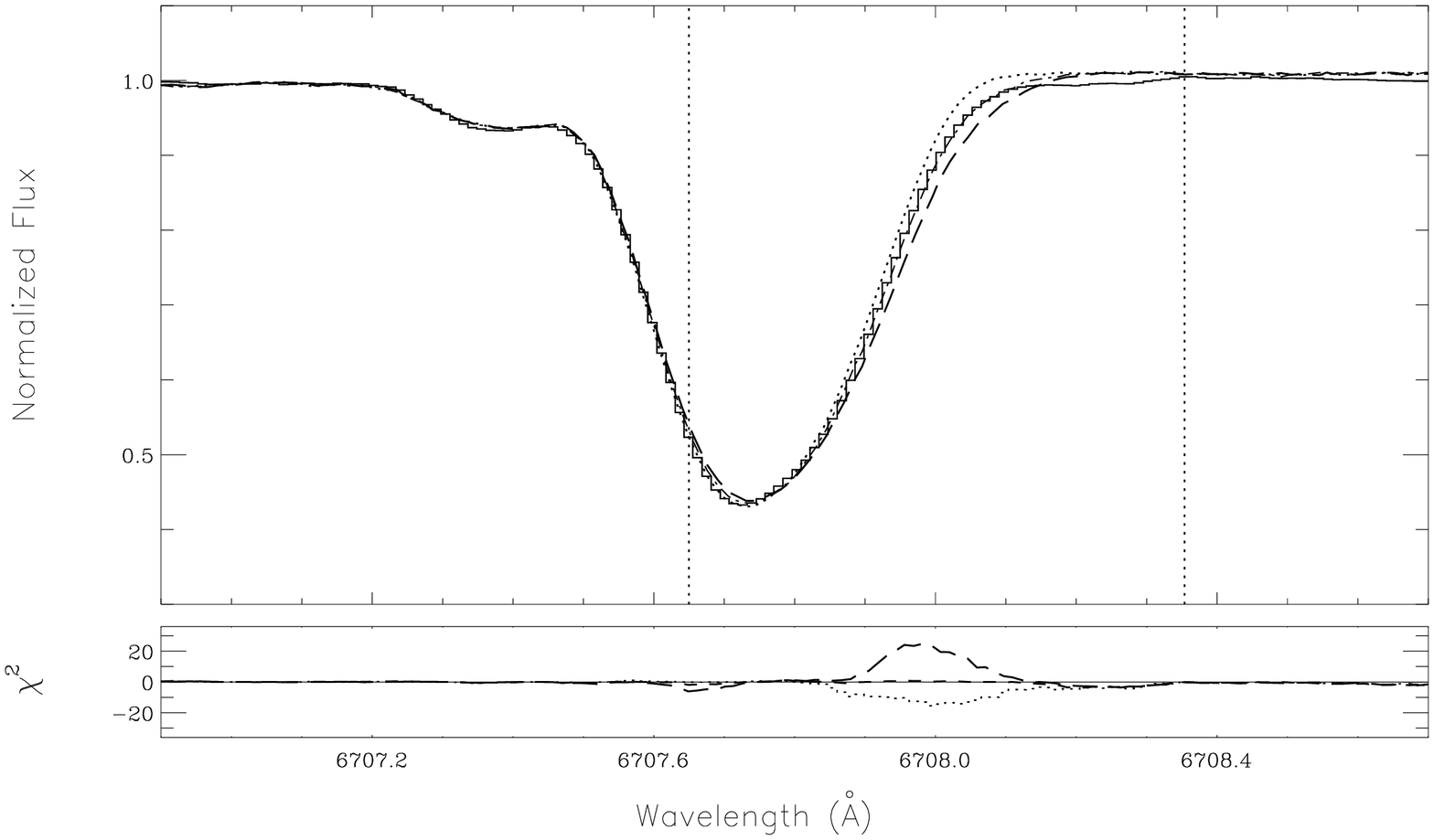}
\includegraphics[scale=0.63,angle=90]{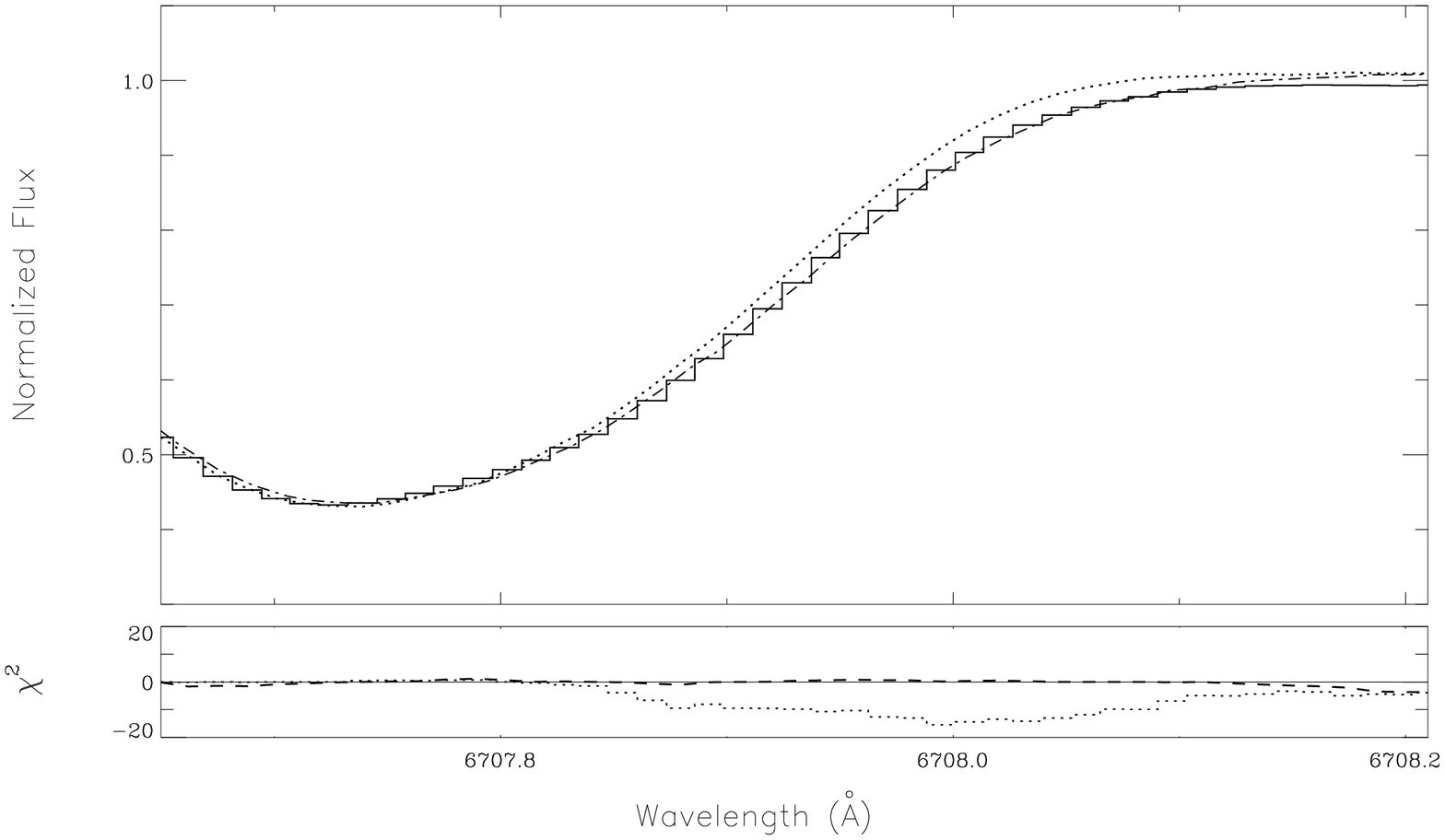}
\caption{
 GJ~117 spectrum for the Li I 6707.8 \AA\ region plotted as a solid histogram.
a. ({\it top-figure})
Over-plotted are PHOENIX models
with $\frac{^{6}Li}{^{7}Li}$ ratios of 0.0 (dotted), 0.05 (dash-dotted), 
and 0.10 (long-dashed) lines. 
The lower panel shows $\chi^2$ residuals from the difference
of the data minus the model (see text).
The interval in which $\chi^2$ was calculated is indicated by the dotted vertical
lines.
b. ({\it bottom-figure}) Shows an expanded view of the lithium line profile with only the
PHOENIX models with $\frac{^{6}Li}{^{7}Li}$ ratios of 0.0 (dotted), 0.05 (dash-dotted)
over-plotted, and their residuals.
}
\label{fig:specmods}
\end{figure*}

We have compared the combined McDonald spectrum to the {\sc Phoenix} models 
using a least square fitting technique where the quality of the fit is determined by $\chi^{2}$ statistics \citep{C05, SLN98}.   

\begin{equation}
\centering
\chi^2_{r} = \frac{1}{dof}\sum^{n}_{1}\frac{(D_i - M_i)^2}{\sigma_i^2}
\end{equation}

where $D_i$ and $M_i$ are the data and model 
fluxes at data point $i$, respectively. 
We defined the variance, $\sigma_i$ as the square root of the counts
at each wavelength. $\sigma$ ranged from $\approx$ 900 in the lithium
line center to $\approx$1350 in the continuum.
The standard continuum normalization was determined using the
{\sc dipso cregs} routine for continuum placement. We performed 
a least-squares fitting to check this continuum level. 
We selected 0.3 \AA\ regions of
continuum to the blue and red side the $^7$Li line profile.  
We fitted the model against the data and adjusting the data by a normalization factor
between 0.98 and 1.02 in 0.001 increments, and 
determined a minimum $\chi^2$ for a normalization factor of 1.003. 
We then used this normalization factor in fitting
each PHOENIX $^6$Li model to the data by varying the $\frac{^{6}Li}{^{7}Li}$ 
fraction in 0.01 increments.  In this way the most probable model 
with $\chi^{2}\sim$1 was determined.
$\chi^{2}$ was computed over the red-side of the $^6$Li + $^7$Li line profile 
in the wavelength range of 6707.65 to 6708.35\AA, giving 54 degrees of freedom (dof).
We show the combined McDonald spectrum and its comparison to 3 models with 
the $\frac{^{6}Li}{^{7}Li}$ fraction ranging from 0.0, 0.05, and 0.10
in Figure~\ref{fig:specmods}a, and an expanded view of the red-side of the lithium line profile
in Figure~\ref{fig:specmods}b. 
The best fitting model had a $\frac{^{6}Li}{^{7}Li}$ = 0.05 with a $\chi^{2}$ of $\approx$58 and the model with no $^6$Li had a $\chi^{2}$ $\approx$ 284.
We show the reduced $\chi^2$ values for each model and Table~3. 
For the residuals Figures~\ref{fig:specmods}ab we include a sign factor 
to $\chi^2$ to indicate whether the observational data was larger than 
(positive) or smaller than (negative) the model.

\begin{figure}
\begin{center}
\includegraphics[scale=0.66, angle=0]{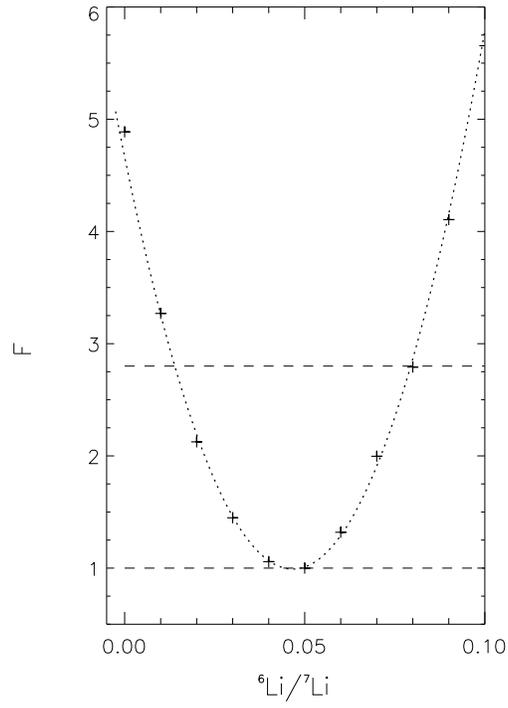}
\end{center}
\caption{
 The F statistic as a function of the $\frac{^{6}Li}{^{7}Li}$ ratio for Li model line profiles as 
compared to the observations for GJ 117. F is defined as the ratio of reduced $\chi^2$ for
each model to the model with the minimum reduced $\chi^2$. The F value for a 90\% confidence
limit ($\approx$2.80) is over-plotted as the dashed horizontal line. A F value of
 1 is also indicated.}
\label{fig:deltachi}
\end{figure}

We computed the F-statistic and compared the reduced $\chi^2$ for
each model to the model with the minimum reduced $\chi^2$ (model with $\frac{^{6}Li}{^{7}Li}$ = 0.05), 
and show this in Figure~\ref{fig:deltachi}.
For 54 degree of freedom the 90\% confidence limit for the F ratio is 
$\approx£$2.8 and this ratio is also shown in Figure~\ref{fig:deltachi}. 
The 1$\sigma$ uncertainty of the $^6$Li fraction is 0.02,
giving $\frac{^{6}Li}{^{7}Li}$ = 0.05$\pm$0.02.


\begin{deluxetable}{lccr}
\label{tab:chisq}
\tablewidth{0pt}
\tablenum{3}
\tablecaption{Results for $\chi^2_r$ for PHOENIX model fitting}
\tablehead{
 \colhead{$\frac{^{6}Li}{^{7}Li}$}
  &\colhead{$\chi^2_r$}
  &\colhead{$\chi^2_r$}
  &\colhead{}
\\
  \colhead{}
  &\colhead{PHX\tablenotemark{a}}
  &\colhead{PHX2\tablenotemark{b}}
  &\colhead{}
}

\startdata
 0.00 &  5.27 & 1.70  \\  %
 0.01 &  3.52 & 1.18 \\  %
 0.02 &  2.29 & 0.98  \\  %
 0.03 &  1.56 & 1.11 \\  %
 0.04 &  1.14 & 1.45  \\  %
 0.05 &  1.08 & 2.06 \\  %
 0.06 &  1.42 & 3.01  \\  %
 0.07 &  2.15 & 4.01 \\  %
 0.08 &  3.01 & 5.36 \\  %
 0.09 &  4.42 & 6.88  \\  %
 0.10 &  6.09 & 8.76  \\  %

\tablenotetext{a}{PHX -- from the PHOENIX line list used in this work.} 
\tablenotetext{b}{PHX2 model with additional Ti I from the line list of 
Reddy et al. 2002}
\enddata
\end{deluxetable}

Recently, \citet{C07} cautioned that the continuum placement and
the uncertainty in the continuum level may dominate the 
errors in the $^6$Li fraction, and 
we tested the effect on the this level on the $^6$Li fraction.
We find that a normalization of 1.01 (1\% increase in the continuum level 
derived a $\frac{^{6}Li}{^{7}Li}$ = 0.03 when re-fitting the lithium line, 
but clearly does not fit the line center (data now
higher than the model)
and this fit has $\chi^2$ about 60 higher than the best fit normalization 
of the continuum, which gives $\frac{^{6}Li}{^{7}Li}$ = 0.05.
Conversely, by lowering the continuum fraction by 1\%, we find a 
$\frac{^{6}Li}{^{7}Li}$ = 0.06, but again this shows an overall poor
fit to the line and continuum and has $\chi^2$ about 100 higher than 
the best fit normalization of the continuum. 

We also investigated if
small wavelength shifts could change the derived $\frac{^{6}Li}{^{7}Li}$
fraction.  Our PHOENIX models are generated with a wavelength increment of 0.01 \AA\
and then re-binned and interpolated onto the observed wavelength
scale. 
Thus any small shifts caused by an uncertainty in the
wavelength scale should be accounted for as the data and model
shift relative to each other. 
However, we tested how small absolute shifts between the data and 
model would change $\chi^2$ and the derived $\frac{^{6}Li}{^{7}Li}$. 
Shifts to the data relative to the model on the order of 
1\% of a resoluton element (0.5 m\AA) only changed $\chi^2$
by a small amount and caused no changes to the computed $\frac{^{6}Li}{^{7}Li}$.
However, we then tested what amount of an absolute shift would cause a 
change of 0.01 in the $\frac{^{6}Li}{^{7}Li}$.
A shift of $+$5m\AA\ (10\% of a resolution
element) is needed to shift the best fitting model for the $\frac{^{6}Li}{^{7}Li}$ 
to 0.06 and similarly a shift of -5 m\AA\ gives the best fitting model to 0.04. 
However, both of these shifts produce a $\chi^2$ approximately 30 higher
than the best fitting model, and such a large shift is not supported
by the wavelength calibration and modelling, and any smaller shifts are 
accounted for in the current uncertainty.
 


We have found a  $\frac{^{6}Li}{^{7}Li}$ of 
0.05$\pm0.02$ for our McDonald observations of GJ~117.
This value is $\approx$3\% higher than our earlier VLT estimate. 
The main difference can be attributed to not including the 
strong Ti I line at 6708.025 \AA. If we include this line at the
strength used by \citet{R02} we find a $\frac{^{6}Li}{^{7}Li}$ of
0.02$\pm$0.01.  The fitting results for this model are also
summarized in Table~3.  Our template star, GJ~211 and other K2
stars in our sample do not show any additional absorption near this
wavelength stronger than 1 m\AA\ and our original PHOENIX model 
appears to be more consistent with GJ~117. An analysis of the 
VLT spectrum with our first PHOENIX model also finds a 
similar result for the $\frac{^{6}Li}{^{7}Li}$
of 0.05$\pm$0.03 and is thus consistent with our higher signal-to-noise
McDonald data.     
The spectra from the two 
epochs 
do not show a significant change above the 
statistical uncertainties and uncertainty in normalizing the spectra.
We note that the VLT spectra have a S/N of 
$\approx$ 400, significantly lower than the McDonald observations. 
%


\subsection{Evidence for flare activity on GJ~117}
The higher rotational velocity and deeper convection zone make GJ~117 
considerably more active than the Sun. In our earlier work \citep{C05} we 
presented arguments for the generation of Li in the atmosphere of GJ~117 
by spallation reactions (for example, see the recent work of \citet{T08}). 
These arguments were based on the strong quiescent 
X-ray luminosity of this source which in turn implied a high time-averaged 
flare energy \citep{DB85}. 

\begin{figure}[t]
\includegraphics[scale=0.66, angle=90]{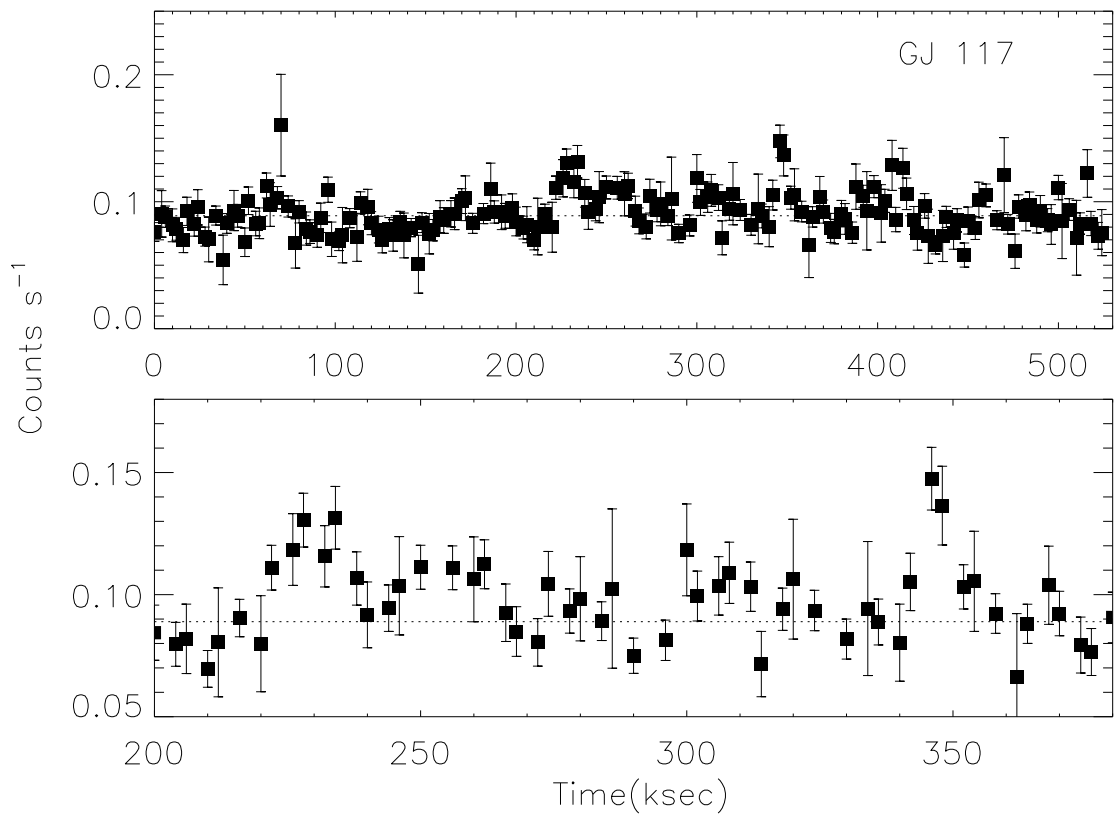}
\caption{EUVE light curve data. The {\it top} panel shows the entire observation
in 2 ksec bins for the Deep Survey 100 \AA\ bandpass. The {\it lower} panel
shows a blow-up of the 200--400 ksec interval showing two flares (see text). 
}
\label{fig:euve}
\end{figure}

We have examined archival EUVE data for GJ~117 for variability. 
A light curve of these 
observations in the EUVE DS 100\AA\ bandpass is shown in Figure~\ref{fig:euve}. The 
observations of GJ~117 shows 3 small flares during the 540 ksec duration. 
The largest flare shows a $\approx$50\% increase over the quiescent rate 
of 0.09 counts s$^{-1}$. 
The variability of GJ~117 was also confirmed found to be significant 
using the Kolmogorov-Smirnov (KS) statistical tests \citep{CM98}.

\section{Concluding Remarks}
\label{conc}
We have obtained 
high resolution McDonald echelle observations of the dK1 star GJ~117 as 
a follow-up to our detection of $^6$Li in this object using
VLT and UVES \citep{C05}.
Here we report the detection of $^6$Li in our McDonald data at the 5\% level 
($\frac{^{6}Li}{^{7}Li}$ = 0.05$\pm0.02$).  
We have used the solar spectrum and template stars to eliminate possible
blends, such as Ti~I, in the $^6$Li spectral region. 
Additionally, bisector analysis showed no significant red asymmetries 
that would affect the lithium line profile.
GJ~117 is much more active than the Sun and 
its X-ray luminosity is at least one order of magnitude higher.
As outlined in \citet{C05}, GJ~117 has the needed energy budget to produce 
the observed 
$^6$Li in spallation reactions on the stellar surface. 
However, we caution that uncertainties in the continuum placement may cause
additional errors that may decrease or increase 
the $^6$Li fraction for the current methods used.
Future high resolution observations during a large flare event on such
Li-rich active dwarf stars could provide definitive proof of spallation
reactions in these stars.

\acknowledgements
We thank David Doss and all of the McDonald staff for their excellent 
support for this project. We thank Dr. David Lambert for
suggested improvements to the manuscript.
We also thank an anonymous referee for suggested improvements.
DC thanks STFC for travel support.
D.J. was  
supported by the projects No 146001 and 146007 financed by Ministry of 
Science of Serbia.


\end{document}